\documentclass[prl,twocolumn,aps,tightenlines,showpacs,floatfix]{revtex4}

\usepackage[dvips]{graphicx}

\usepackage{dcolumn}     % Align table columns on decimal point

\renewcommand{\case}{\frac}
\newcommand{\boldtau}{\mbox{\boldmath$\tau$}}
\newcommand{\boldsigma}{\mbox{\boldmath$\sigma$}}

\begin{document}
 
{  % \tighten

%%% DRAFT --- DRAFT --- DRAFT --- DRAFT --- DRAFT --- DRAFT --- DRAFT --- DRAFT --- DRAFT

\title{Can Modern Nuclear Hamiltonians Tolerate a Bound Tetraneutron?}

\author
{ Steven C. Pieper\cite{scp} }

\affiliation
{Physics Division, Argonne National Laboratory, Argonne, Illinois 60439}

\date{\today}

\begin{abstract}
I show that it does not seem possible to change modern nuclear Hamiltonians 
to bind a tetraneutron without destroying many other successful predictions
of those Hamiltonians.  This means that, should a recent experimental claim
of a bound tetraneutron be confirmed, our understanding of nuclear forces
will have to be significantly changed.  I also point out some errors
in previous theoretical studies of this problem.
\end{abstract}
 
\pacs{13.75.Cs, 21.30.-x, 21.45.+v, 21.60.Ka, 21.65.+f}

\maketitle

}
 
%  \section{INTRODUCTION}

An experimental claim of the existence of a bound tetraneutron cluster
($^4$n) was made last year~\cite{4n-exp}.  Since then a number of
theoretical attempts to obtain such bound systems have been made, with
the conclusion that nuclear potentials do not bind four
neutrons~\cite{Tim02,Ber,Tim03}.  However these studies have been made
with simplified Hamiltonians and only approximate solutions of the
four-neutron problem.  In this paper I use modern realistic nuclear
Hamiltonians that provide a good description of nuclei up to $A=10$ and
accurate Green's function Monte Carlo (GFMC) calculations to improve
this situation.  (A list of earlier experimental and theoretical
studies, also with generally negative conclusions, may be found in
~\cite{Tim02,Ber,Tim03}.)

A series of papers \cite{PPCPW97,WPCP00,PVW02}, have presented the
development of GFMC for calculations of light nuclei (so far up to
$A$~=~10) using realistic two-nucleon ($N\!N$) and three-nucleon
($N\!N\!N$) potentials.  For a given Hamiltonian, the method obtains
ground and low-lying excited state energies with an accuracy of 1---2\%.
I use this method in the present study; tests similar to those reported
in the above papers have verified that the energies reported here have
similar accuracies, with two exceptions: 1) when the energies are very
close to 0, the error is probably a few 100 keV; and 2) the $^4$H
calculations contain a technical difficulty that might be introducing
systematic errors of up to 1~MeV.  The reader is refered to~\cite{PW01}
for a review of the nuclear GFMC method or to the previously cited
papers for complete details of how the present calculations were made.

By using the Argonne $v_{18}$ $N\!N$ potential (AV18)~\cite{WSS95} and
including two- and three-pion exchange $N\!N\!N$ potentials, a series of model
Hamiltonians (the Illinois models) were constructed that reproduce energies
for $A$~=~3---10 nuclei with rms errors of 0.6---1.0~MeV~\cite{PPWC01}.  The
best model, the AV18+Illinois-2 (AV18/IL2) model, is used in the present study.

GFMC starts with a trial wave function, $\Psi_T$, which determines the
quantum numbers of the state being computed. For $p$-shell nuclei studied
in the above references, the Jastrow part of $\Psi_T$ contains four nucleons with an
alpha-particle wave function and $A-4$ nucleons in $p$-shell orbitals.
This is multiplied by a product of non-central two- and three-particle
correlation operators.
I use $\Psi_T$ for $^4$n with the same structure except there are two neutrons
in a $^1\!S_0$ configuration and two in the $p$-shell.  The total
$J^\pi$ of the $^4$n ground state is assumed to be $0^+$.  There are two
possible symmetry states in the $p$-shell using $LS$ coupling: $^1$S[22]
and $^3$P[211]; both are used in these calculations.  I could find no
$\Psi_T$ that gave a negative energy for $^4$n using the AV18/IL2 model.
GFMC calculations, using propagation to very large imaginary time
($\tau=1.6$~MeV$^{-1}$), also produced positive energies that steadily
decreased as the rms radius of the system increased.

\begin{figure}[ht!]
\centering
\includegraphics[width=2.5in,angle=270]{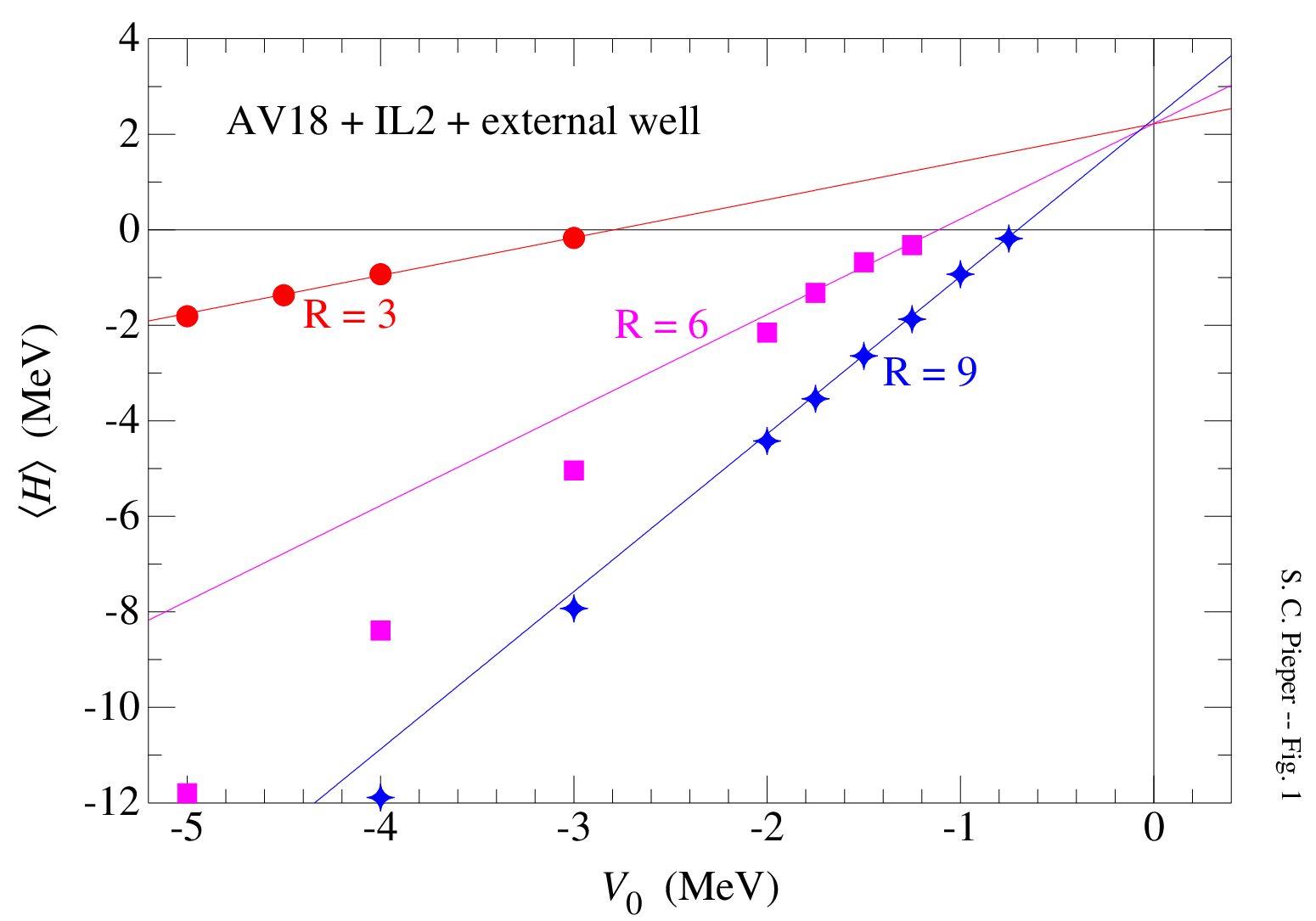}
\caption{Energies of $^4$n in external wells versus the well-depth parameter, $V_0$.}
\label{fig:4n_ws_extrap}
\end{figure}

In a second study, I added artificial external wells of Woods-Saxon shape to
the AV18/IL2 Hamiltonian and used GFMC to find the resulting total energies of
the four neutrons.  Figure~\ref{fig:4n_ws_extrap} shows results for wells
with radii $R=3$, 6, and 9 fm (all have diffuseness parameters of 0.65 fm)
and varying depth parameter, $V_0$.  It seems clear that four neutrons
become unbound (have positive energy) significantly before the well depth
is reduced to zero.  Linear fits to the least bound energies for each Woods Saxon 
radius parameter are also shown; these extrapolate to an energy of +2 MeV when the
external well is removed.  (These least-bound solutions have large rms radii.
A transition from the indicated linear
behavior to a steeper linear behavior is observed for deeper wells; this transition
is associated with a change to much smaller rms radii solutions.  The steeper fits
of course extrapolate to much larger positive energies.)
This suggests that there might be a $^4$n resonance near 2 MeV, but since
the GFMC calculation with no external well shows no indication of
stabilizing at that energy, the resonance, if it exists at all, must be
very broad.  In any case, the AV18/IL2 model does not produce a bound $^4$n.

The authors of~\cite{4n-exp} suggest that only small modifications
of existing nuclear Hamiltonians may be necessary to bind four neutrons.
To study this possibility, I made a number of modifications to the AV18/IL2 model.
In each case the modification was adjusted to bind $^4$n with
an energy of approximately $-0.5$~MeV; the consequences of this
modification for other nuclei were then computed.  Four of the modifications
are reported here:  long- and moderate-range changes of the 
$N\!N$ potential in the $^1\!S_0$ partial wave; introduction of
an additional $N\!N\!N$ potential that acts only in total isospin $T=\case{3}{2}$
triples; and introduction of a $N\!N\!N\!N$ potential that acts only
in $T=2$ quadruples.  In all cases the complete AV18/IL2 Hamiltonian was used
with the additional term.

The strong-interaction part of the AV18 $N\!N$ potential consists of
one-pion exchange with the generally accepted value of $f_\pi^2/4\pi=0.075$,
moderate-range terms that are associated with two-pion exchange
but which have phenomenologically adjusted strengths, and a short-ranged
completely phenomenological part.  The potential is written in terms
of operators which can be used to produce the potential for any
partial wave.  By making correlated changes to the radial parts
of the central, $\boldsigma_{i}\cdot\boldsigma_{j}$,
$\boldtau_{i}\cdot\boldtau_{j}$, and 
$\boldsigma_{i}\cdot\boldsigma_{j} \boldtau_{i}\cdot\boldtau_{j}$
operators, one can change the potential in just the $S=0, T=1$
partial waves: $^1\!S_0, ^1\!D_2, ^1\!G_4$, etc.  A corresponding
change to the four ${\bf L}^2$ operator terms can cancel the
change in the $^1\!D_2$ partial wave, leaving a change to basically
only the $^1\!S_0$ partial wave.  

\begin{figure}[ht!]
\centering
\includegraphics[width=3.5in]{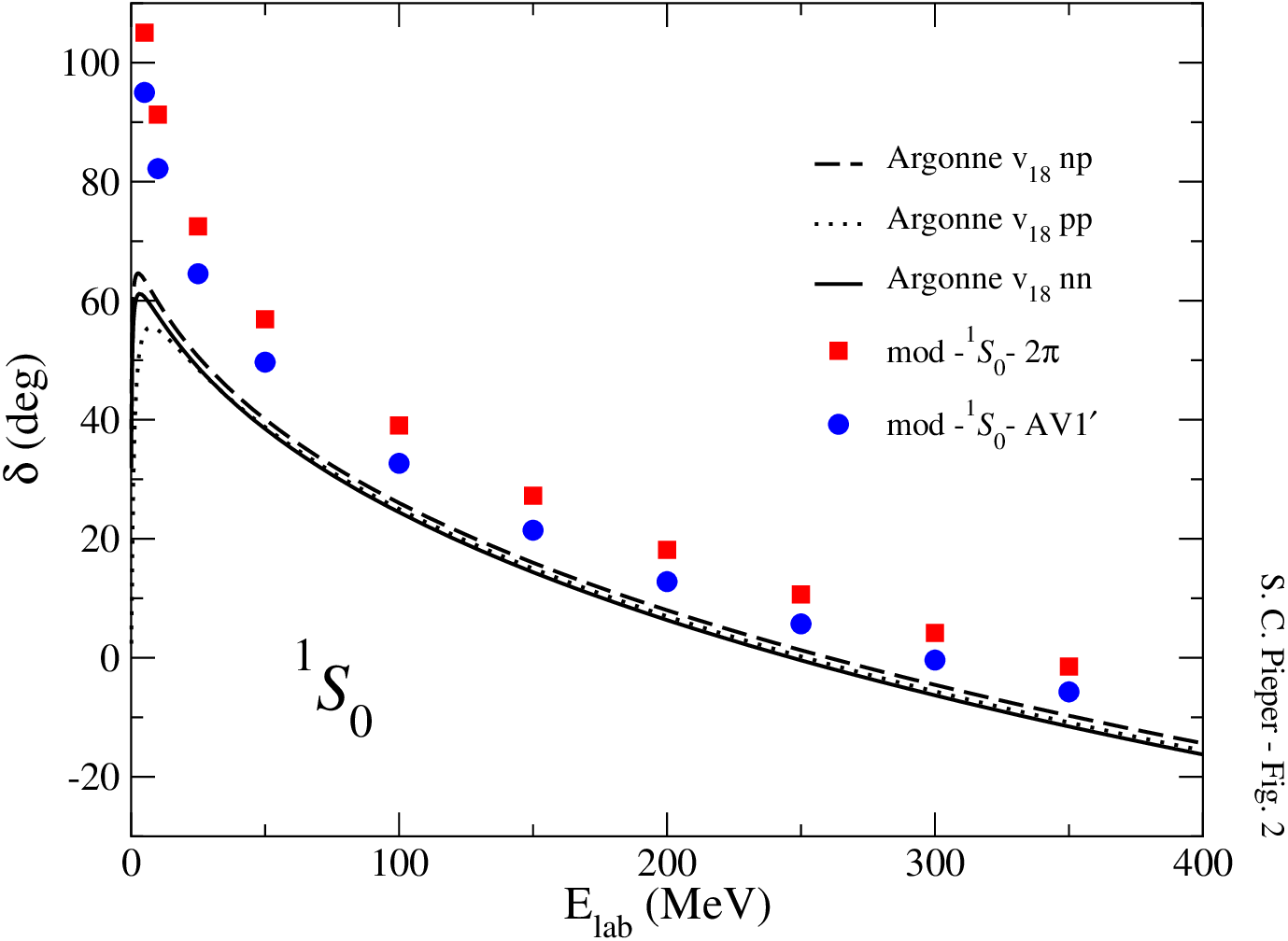}
\caption{$^1\!S_0$ phase shifts from AV18 and modifications to it.
The lines show the pp, pn, and nn phase shifts for the unmodified
AV18 while the symbols show the modified results.}
\label{fig:1s0_phase}
\end{figure}

In the first such modification of the AV18, I changed just the two-pion
range part of the $^1\!S_0$ partial wave, so as to leave the
theoretically well established one-pion part unaffected.  Increasing
this two-pion strength by 4.9\% results in a $^4$n energy of
$-0.87$(3)~MeV.  (The statistical errors in Monte Carlo computed numbers
are shown in parentheses only when they exceed unity in the last quoted
digit.)  As is shown by the points labeled ``mod-$^1\!S_0$-$2\pi$'' in
Fig.~\ref{fig:1s0_phase}, this changes the $^1\!S_0$ phase shifts by
12$^\circ$ over a large energy range and produces a bound dineutron
(the energy is $-0.88$~MeV).  These changes far exceed those allowed by
modern phase shift analysis.  A somewhat smaller change that produces a
bound $^4$n can be made by using the AV1$^\prime$ potential~\cite{WP02}
in the $^1\!S_0$ partial wave (and AV18 in the other partial waves);
this results in a $^4$n energy of $-0.52$~MeV and about a 50\% smaller
change in the $^1\!S_0$ phase shifts (the points labeled
``mod-$^1\!S_0$-AV1$^\prime$'' in the figure).  However again $^2$n is
bound, this time with an energy of $-0.42$~MeV.  Note that the
one-pion-range part of the potential is also changed in
mod-$^1\!S_0$-AV1$^\prime$.  These $^2$n and $^4$n bound states are
quite diffuse; the rms radii are respectively 2.8 and 3.6~fm for the two
$^2$n cases and 7.3 and 10.3~fm for the $^4$n.  The $^4$n pair
distributions have a peak containing about two pairs with a structure
close to that of the $^2$n pair distribution and a long tail.  Thus the
$^4$n looks like two widely separated dineutrons.

\begin{figure}[ht!]
\centering
\includegraphics[width=2.35in,angle=270]{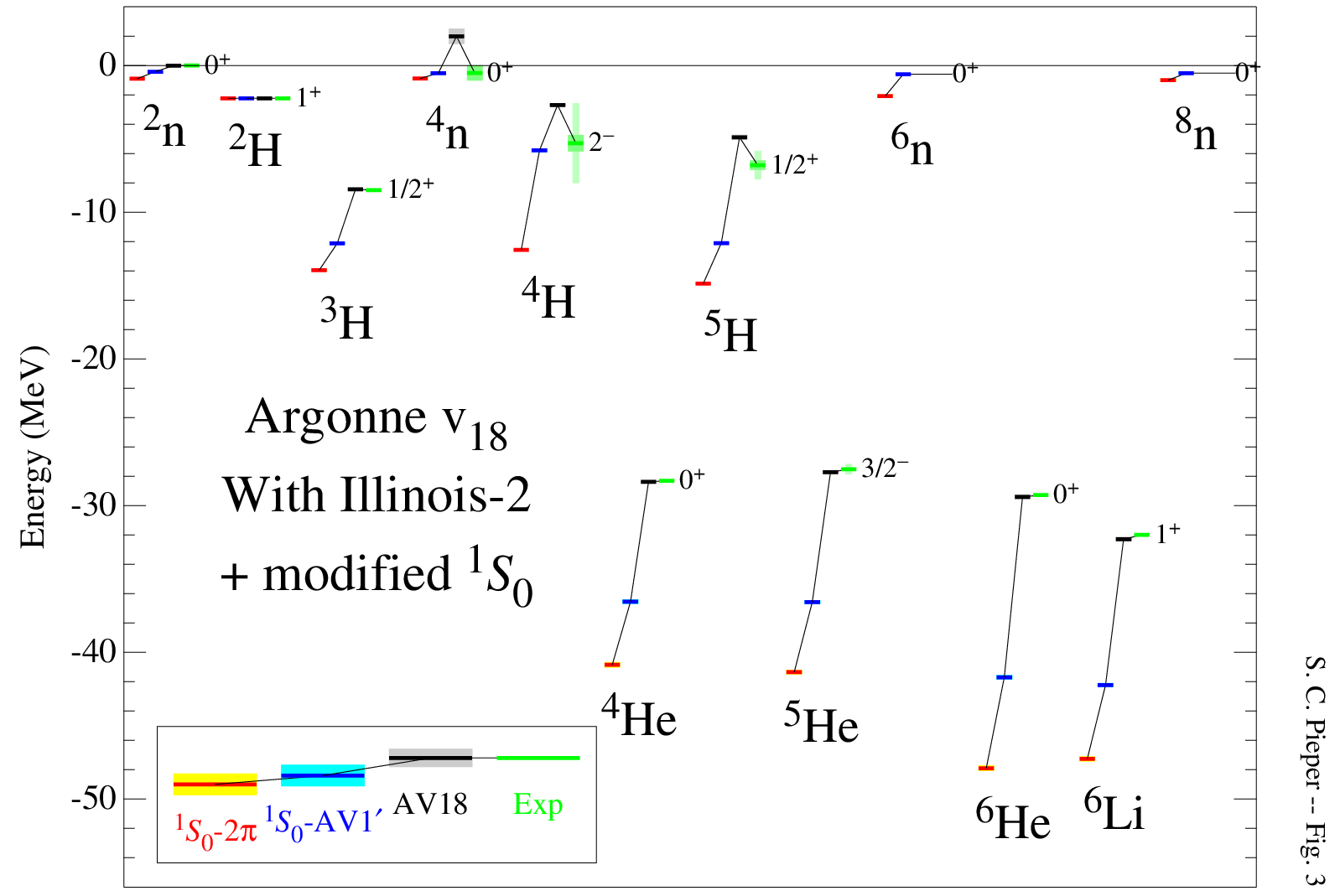}
\caption{Energies of nuclei and neutron clusters computed with the
AV18/IL2 Hamiltonian with modified 
$N\!N$ potentials ($^1\!S_0$-$2\pi$ and $^1\!S_0$-AV1$^\prime$) and
with no modification (AV18), compared with experimental values for known nuclei.}
\label{fig:2n_8n-2b}
\end{figure}

As noted, these modifications of the $^1\!S_0$ potential to produce
minimally bound tetraneutrons also produce dineutrons with about the
same binding energies; thus they are physically unacceptable
modifications.  Figure~\ref{fig:2n_8n-2b} shows that they also introduce
large changes to the binding energies of other nuclei; for example $^3$H
is $\sim$50\% overbound and $^5$H is stable or almost stable against
breakup into $^3$H+n+n as opposed to being a resonance in that
channel~\cite{5Hexp}.  Also six and eight neutrons form bound systems, 
although three and five do not..

The authors of Refs.~\cite{Tim02,Ber,Tim03} concluded that the non-realistic
Volkov potentials~\cite{Volkov} do not bind $^4$n.  However these
potentials do have bound dineutrons and the above results suggest
that they thus might bind $^4$n.  To study this I made calculations
using the first four Volkov potentials in all partial waves and
no $N\!N\!N$ potential.  These potentials indeed do bind $^4$n
with energies of $-0.91$,  $-1.04$, $-0.47$ and $-0.71$~MeV while the 
$^2$n energies are $-0.56$, $-0.60$, $-0.35$, and $-0.42$~MeV, respectively.
The rms radii of the $^4$n systems are all about 11.5~fm, which
may explain why these bound states were not discovered in Refs.~\cite{Tim02,Ber,Tim03}.
The variational energies for $^4$n with modifications to the AV18/IL2 Hamiltonian
are positive; that is, only with GFMC improvement does the system become bound.
However for the simpler Volkov potentials, the $\Psi_T$ already
give negative energies and the GFMC just improves these energies.

It must be emphasized that these bound $^4$n results do not at all
support an experimentally bound $^4$n. The more than 35-year old
Volkov potentials are not realistic;  they produce bound $^2$n,
with the same binding energies as their deuterons; they have no
tensor or $L{\cdot}S$ terms; and they can not reproduce modern phase shift
analyzes in any partial wave.  The one thing in their favor is that,
by having a space-exchange component, they introduce some saturation
in $p$-shell nuclear binding energies; however with just one radial
form they are even simpler than the space-exchange AVX$^\prime$ introduced
in Ref~\cite{WP02}.

The above results show that it is not possible to bind $^4$n by
modifying the $^1\!S_0$ potential without severely disrupting other
nuclear properties.  The next $N\!N$ possibility is the $^3\!P_J$
channel.  The net effect of these is a small repulsion in neutron
systems.  Setting this term to zero had very little effect on
$^4$n; one would have to introduce significant attraction to
bind $^4$n and then again many other nuclear properties would
be unrealistically changed.

It has been suggested that modifications to the $N\!N\!N$ or
$N\!N\!N\!N$ potentials, which are experimentally much less constrained
than the $N\!N$ potential, could be used to bind $^4$n.  Timofeyuk added
a central $N\!N\!N\!N$ potential to bind $^4$n, but found that it
resulted in $^4$He being bound by about 100~MeV~\cite{Tim02,Tim03}.
However, as she suggests, one should try less disruptive things.  A
$N\!N\!N$ potential that acts only in $T=\case{3}{2}$ triples would have
the same effect on $^4$n as one with no isospin dependence, but no
effect on $^3$H and $^4$He because the contain only $T=\case{1}{2}$
triples.  A $N\!N\!N\!N$ $T=2$ potential would also not effect $^5$He
and $^6$Li.

To study such possibilities, I added potentials of the forms
\begin{eqnarray*}
V_{ijk}(T\!=\!\textstyle{3\over2}) &=& V_3 \sum_{cyclic}[ Y(r_{ij})Y(r_{jk}) ] 
P(T\!=\!\textstyle{3\over2}) , \\
V_{ijkl}(T\!=\!2) &=& V_4 \sum_{cyclic}[ Y(r_{ij})Y(r_{jk})Y(r_{kl}) ] P(T\!=\!2) , \\
Y(r) &=& \frac{e^{-m_\pi r}}{m_\pi r} [ 1 - e^{-(m_\pi r)^2} ]^2 , 
\end{eqnarray*}
to the AV18/IL2 Hamiltonian.
Here $m_\pi$ is the pion mass, the $P$ are projectors onto the indicated
isospin states, and $V_3$ and $V_4$ were chosen to produce $^4$n with
$\sim -0.5$~MeV energy.  These forms have the longest range
that is possible from strong interactions; the cutoff makes the
radial forms peak at 1.55~fm.  Using more confined radial
forms only increases the problems reported below.  

It turns out that the couplings must be quite large to produce the
minimally bound $^4$n: $V_3=-440$ and $V_4=-4750$~MeV, which result in
$^4$n energies of -0.60(5) and -0.55(6)~MeV.  This
can be understood as follows.  If the $N\!N$ potential is used to bind
$^4$n, the pairs can sequentially come close enough to feel the
attraction; this allows the four neutrons to be in a diffuse, large
radius, distribution.  However a $N\!N\!N$ potential requires three
neutrons to simultaneously be relatively close and thus the density of
the system must be much higher.  Indeed the rms radii of the $^4$n for
the $V_{ijk}(T\!=\!\textstyle{3\over2})$ case is only 1.88~fm, while
that for $V_{ijkl}(T\!=\!2)$ is 1.61~fm.  Such small radii result in
kinetic energies that are an order of magnitude more than those for the
$^4$n systems bound by modified $^1\!S_0$ potentials; for the
$V_{ijk}(T\!=\!\textstyle{3\over2})$ case the expectation value of the
kinetic energy is $\sim$87~MeV, while those of the $N\!N$ and $N\!N\!N$
potentials are -49 and -38~MeV, respectively. (As is discussed in
Ref.~\cite{PPCPW97}, GFMC directly computes only $\langle H \rangle$;
other expectation values involve extrapolations.  Here I have reported
the extrapolated potential values and subtracted these from $\langle H
\rangle$ to get the kinetic energy.)

\begin{figure}[ht!]
\centering
\includegraphics[width=2.35in,angle=270]{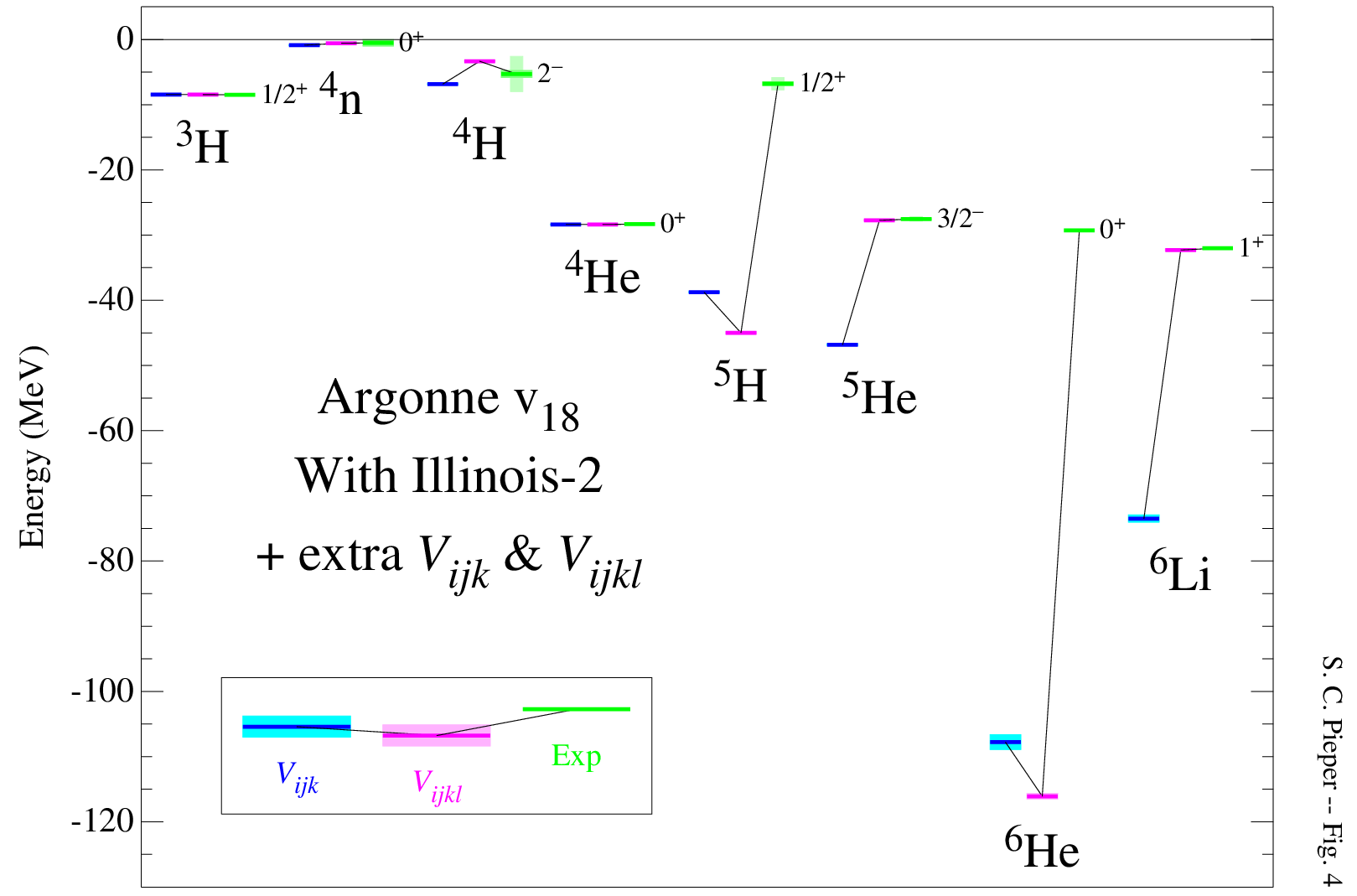}
\caption{Energies of nuclei and neutron clusters computed with modified $N\!N\!N$ 
and $N\!N\!N\!N$ potentials.}
\label{fig:2n_8n-3b4b}
\end{figure}

The very large coupling constants for the
$V_{ijk}(T\!=\!\textstyle{3\over2})$ and $V_{ijkl}(T\!=\!2)$ potentials
mean that they have a large, even catastrophic, effect on any nuclear
system in which they can act.  This is shown in
Fig.~\ref{fig:2n_8n-3b4b}; for example
$V_{ijk}(T\!=\!\textstyle{3\over2})$ doubles the binding energy of
$^6$Li and triples that of $^6$He, while $V_{ijkl}(T\!=\!2)$, which can
have no effect on $^6$Li, quadruples the binding energy of $^6$He.  As
noted before both of these potentials have no effect on $^4$He.  Both
potentials make $^5$H stable by more than 25~MeV against $^3$H+n+n.
However the most dramatic result of these potentials is that every
investigated pure neutron system with $A>4$ is extremely bound and in
fact is the most stable ``nucleus'' of that A.  For
$V_{ijk}(T\!=\!\textstyle{3\over2})$ the energies are -62, -220, and
-650~MeV respectively for $A=5,6,8$, while for $V_{ijkl}(T\!=\!2)$ they
are -358, -1370, and -6690~MeV.

These enormous bindings indicate that matter will collapse with such
potentials.  This is to be expected for purely attractive many-nucleon
potentials.  One should add a shorter-ranged stronger repulsion to
obtain saturation.  Such a repulsion might improve the results for
$A\geq6$ nuclei.  I attempted to study this by using a repulsive term
with Yukawa radial forms of range $2m_\pi$.  However, in order to get
any appreciable effect on $^6$He, the repulsive coupling has to be made
quite large; this then requires at least a doubling of the attraction to
still bind $^4$n; this results in potentials that are so strong that the
GFMC starts to become unreliable.  The apparent impossibility of
correcting the $A=6$ results by such a term may also be seen from the
rms radii of the $^4$n reported above; they are smaller than the
experimental value for $^6$Li and reasonable $^6$He radii.  Thus a
short-ranged repulsion that still leaves the $^4$n bound will certainly
result in $A=6$ nuclei with too small rms radii.

In all of these cases I have made isospin-conserving modifications to
the AV18/IL2 Hamiltonian; thus there have been $T=1$ additions to the
$N\!N$ potential, or a $T=\case{3}{2}$ addition to the $N\!N\!N$
potential, or a $T=2$ addition to the $N\!N\!N\!N$ potential.  It might
be objected that the modifications should have been made only for nn
pairs or nnn triples or nnnn quadruples since the nuclear force is least
well determined for such systems.  Such changes would mean much larger
charge-symmetry breaking and charge-independence breaking potentials
than are presently accepted.  But even so, the changes to the $N\!N$
force, if limited to just nn pairs, would still bind two neutrons which
would change the experimental scattering length from $\sim-18$~fm to a
positive value.  Such a nn potential would still bind $^6$n and $^8$n.
I estimate that it would still increase the binding of $^3$H by 3 MeV
while it would have no effect on $^3$He.  Thus the Nolen-Schiffer energy
for the $A=3$ system would be some five times too large.  Many of the
devastating effects shown in Fig.~\ref{fig:2n_8n-3b4b} would similarly
persist even if the potentials were limited to nnn triples or nnnn
quadruples.

In conclusion, should the results of Ref.~\cite{4n-exp} be confirmed,
our current very successful understanding of nuclear forces would
have to be severely modified in ways that, at least to me, are not at 
all obvious.

\acknowledgments
I thank R. B. Wiringa for many useful suggestions during the course of this
work, especially concerning modifications to the AV18 potential; I also
thank V. R. Pandharipande for a critical reading of an early version
of the manuscript.
The calculations were made possible by generous grants of time on the
Chiba City Linux cluster of the Mathematics and Computer Science
Division, and friendly-user access to the Jazz Linux cluster of the
Laboratory Computing Resource Center, both at Argonne National
Laboratory.  This work is supported by the U. S. Department of Energy,
Nuclear Physics Division, under contract No. W-31-109-ENG-38.

\vfill

\end{document}